\documentstyle[12pt]{article}
\newcommand{\eq}[1]{(\ref{#1})}
\newcommand{\diff}{\partial}
\newcommand{\beq}{\begin{equation}}
\newcommand{\eeq}{\end{equation}}
\newcommand{\beqn}{\begin{eqnarray}}
\newcommand{\eeqn}{\end{eqnarray}}
\newcommand{\cD}{{\cal D}}
\def\cD{{\cal D}}
\def\cN{{\cal N}}

\title{\hfill{\small ITEP-TH-30/01}
\\~~\\ On a Modification of the Boundary State Formalism
in Off-shell String Theory }

\author{E. T. Akhmedov$^{~1,2}$, M. Laidlaw$^{~1}$ and  G. W. Semenoff$^{~1}$
\\~~\\
1. Department of Physics and Astronomy, \\
 University of British Columbia, \\
Vancouver, British Columbia, Canada V6T 1Z1.
\\~~\\
2. Institute of Theoretical and Experimental Physics,\\
B. Cheremushkinskaya 25, 117259\\ Moscow, Russia}

\begin{document}


\maketitle






\abstract{ We examine the application of boundary states in computing
amplitudes in off-shell open string theory.  We find a straightforward
generalization of boundary state which produces the correct matrix
elements with on-shell closed string states.}

\newpage

\noindent
{\bf 1.}Background independent string field
theory~\cite{Witten:1992qy}-\cite{Shatashvili} is an
interesting approach to the problem of defining string theory
off-shell.  It has recently received a lot of attention, particularly
as an approach to understanding the properties of unstable
D-branes~\cite{Gerasimov:2000zp}-\cite{Kutasov:2000qp}.

A concrete problem in the study of off-shell string theory in this
formalism has been to understand its behavior in the a background
tachyon field which is a quadratic function of the coordinates.  This
gives a tractable model where one can study phenomena such as tachyon
condensation and D-brane-anti-D-brane
annihilation~\cite{Gerasimov:2000zp}-\cite{Arutyunov:2001nz}.  In this
context, the concept of boundary state, which describes the coupling
of string world-sheets to a D-brane has been used by several
authors~\cite{deAlwis:2001hi}-\cite{Fujii:2001qp}.

The boundary state formalism could be useful in many circumstances:
in computing D-brane tensions and
cylinder amplitudes as well as in looking for the gravity counterparts
of D-branes~\cite{DiVecchia:1999fx,DiVecchia:1999rh}.
This formalism was originally used to factorize open string amplitudes in terms of
closed string states.  This could be valuable in understanding the
relationship between closed and open strings which is one of the
central problems in uncovering the underlying symmetry of string
theory.  In the operator approach to string perturbation theory, the
boundary state contains the coupling of closed strings to a D-brane.

In this Letter, we shall suggest a generalization of boundary states
to be applied to the problem of computing off-shell amplitudes of the
open Bosonic string.  We consider the boundary state for a D-brane
with a tachyon condensate and take the special case where the tachyon
has a quadratic profile.

Then, we will examine the coupling of massless closed string states to
the boundary state.  There are two ways of analyzing this coupling.
The first uses a sigma model approach.  In that case, we insert the
vertex operator for a graviton into the sigma model path integral with
disc geometry and compute the expectation value.  In the second
approach, we construct a boundary state for the D-brane with tachyon
condensate and consider the inner product of this state with the
on-shell closed string graviton.  We find that the result does not
agree with the sigma model computation.

We then explain the reason for this disagreement and invent a modified
boundary state which has the property that its inner products with all
massless closed string states agree with amplitudes computed by
inserting vertex operators for massless states into the sigma model
path integral.

\noindent
{\bf 2.}First, consider the sigma model which defines background
independent string field theory.  We will use the functional integral
representation of the partition function of the Bosonic string.  The
world-sheet is the unit disc and the target space is 26-dimensional
Euclidean space.  The Bosonic string action is supplemented by a
boundary term which contains the quadratic open string tachyon
background:

\beqn\label{gaus}
Z(g,F,T_0,U) = (2\pi\alpha')^{13} \, \int \cD X \, e^{- S(X, g, F, T_0, U)}  \nonumber \\
= (2\pi\alpha')^{13} \, \int \cD X \exp\left\{ - \frac12 \int_D d^2\sigma \, g_{\mu\nu} \,
\diff^a X^\mu (\sigma) \, \diff^a X^\nu(\sigma)  - \right. \nonumber \\ \left. - \oint_{\diff D} d \theta
\left(\pi \alpha' F_{\mu\nu} X^\mu (\theta) \diff_\theta X^\nu (\theta) +
\frac{1}{2\pi}T_0 + \frac{\alpha'}{4} U_{\mu\nu} \, X^\mu(\theta) \, X^\nu(\theta)\right) \right\},
\eeqn
Here, $\alpha'$ is the inverse string tension.  The world-sheet is a
disc, $D$, for which we shall use complex coordinates
$z=e^{-\sigma_1 - {\rm i} \, \sigma_2}$ with $0 < \sigma_1 < \infty$ and $0 \le
\sigma_2 \le 2\pi$ is the parameterization of the disc $D$ (or of the
infinitely long half cylinder). We shall also sometimes use the
coordinate $z=\rho e^{-{\rm i} \, \theta}$.  $0\le\theta\le 2\pi$ is the
parameterization of the boundary of the disc $\diff D$.  $X^\mu
(\sigma), \, \, \mu=1,...,26$ are maps of the string into the target
space with the constant metric $g_{\mu\nu}$, $T(X) = T_0 +
\frac{\pi \, \alpha'}{2} U_{\mu\nu} \, X^\mu \, X^\nu$ is the tachyon
profile with the constant $T_0$ and constant matrix $U_{\mu\nu}$ of
some rank and $F_{\mu\nu}$ is constant gauge field strength.
This functional integral is taken with boundary conditions:

\beq\label{boundary}
g_{\mu\nu} \, \diff_n X^\nu(\theta) + 2\pi\alpha' F_{\mu\nu} \diff_t X^\nu(\theta) + \frac{\alpha'}{2} U_{\mu\nu} X^\nu(\theta) = 0,
\eeq
where $\diff_n$ and $\diff_t$ are the normal and tangential derivatives to the boundary $\diff D$. We
use a non-standard normalization of $X$ as in \cite{Fradkin:1985qd}.
It is related to the standard one via the rescaling by
$\sqrt{2\pi\alpha'}$.

The theory \eq{gaus} is not conformaly invariant and represents a
special example of the background independent string field
theory~\cite{Witten:1992qy}-\cite{Shatashvili:1993kk}.
Because of the conformal anomaly, this theory explicitly depends on
the conformal factor of the world-sheet metric. The convention is to
consider the theory on the unit disc with flat metric. The main
advantage of the case of \eq{gaus} is that the theory is gaussian and
therefore is exactly
solvable~\cite{Witten:1992qy}-\cite{Witten:1993ed},\cite{Kutasov:2000qp}-\cite{Gerasimov:2001pg}.
For example, the renormalization group flow of the parameters
$U_{\mu\nu}$ in the functional integral \eq{gaus} describes the
annihilation of a D25-brane in Bosonic string
theory~\cite{Kutasov:2000qp}. If the rank of the initial matrix
$U_{\mu\nu}$ is $26-p$, what is left after the annihilation of the
D25-brane is a D$p$-brane.  This arises from the fact that the
$\beta$-function for $U$ is $\beta_U = - U$ and, hence, $U$ flows to
zero in the ultraviolet and to infinity in the infrared limits. Thus,
we see from \eq{boundary} that (if $F=0$) the Neumann boundary
conditions present in the UV limit where $U\sim 0$ evolve to Dirichlet
boundary conditions on $26-p$ coordinates, with the rest of the
coordinates still obeying Newman boundary conditions.  These final boundary
conditions describe a Dp-brane.

The  functional integral \eq{gaus} is readily computed
\cite{Witten:1992qy}-\cite{Witten:1993ed}:

\beqn\label{result}
Z(g,F,T_0,U) = \nonumber \\ = (2\pi\alpha')^{13} \, e^{- T_0} \, \prod^{\infty}_{m=1}
\frac{1}{\det\left(g + 2\pi\alpha' F + \frac{\alpha'}{2} \frac{U}{m}\right)} \, \int d
x_0 \, e^{- \pi \alpha' \frac{U_{\mu\nu}}{2} \, x_0^\mu \, x_0^\nu} =
\nonumber \\ = (2\pi\alpha')^{13} \, \frac{1}{\sqrt{\pi
\alpha'\det(U)}}\, e^{- T_0} \,
\prod^{\infty}_{m=1} \frac{1}{\det\left(g + 2\pi\alpha' F + \frac{\alpha'}{2}
\frac{U}{m}\right)},
\eeqn
where $x_0$ is the zero mode of $X$ and the determinant is taken over
the $\mu$ and $\nu$ indexes.

The expression \eq{result} is divergent, using $\zeta$-function
regularization \cite{Kraus:2001nj} one finds:

\beq\label{div}
Z(g,F,T_0,U) \propto e^{-T_0} \, \sqrt{\det\left(\frac{g + 2\pi\alpha' F}{\pi\alpha'U}\right)}
\det\Gamma\left(1 + \frac{\alpha'U/2}{g + 2\pi\alpha' F}\right),
\eeq
where $\Gamma(g)$ is the $\Gamma$-function.  The dependence of the
transcendental functions on the matrix $U$ is assumed to be defined by
their Taylor expansion.  The divergence in \eq{div} as $U\to 0$ is due
to the infinite volume of the D-brane (and becomes a volume factor in
that limit).

We would like to consider interactions of the D25-brane \eq{gaus} with
massless closed string fields.  For example the D25-brane tension can
be extracted from the expectation value of the graviton vertex
operator.  Consider the correlator:

\beq\label{corr}
\left\langle \int_D d^2\sigma h_{\mu\nu} \diff^a X^\mu(\sigma) \diff^a X^\nu (\sigma) \right\rangle_{F,T_0,U},
\eeq
where the averaging taken in the functional integral \eq{gaus}.
$h_{\mu\nu}$ is a constant traceless matrix which defines the
polarization of the graviton, and we could consider in exactly the same
manner
correlators corresponding to the anti-symmetric tensor field $B$ or
to the dilaton.

It is easy to see that \eq{corr} is given by:

\beqn\label{answer}
\left\langle \int_D d^2\sigma h_{\mu\nu}
\diff^a X^\mu(\sigma) \diff^a X^\nu (\sigma) \right\rangle_{F,T_0,U} =
h_{\mu\nu} \,\left(\frac{\delta Z(g+h',F,T_0,U)}{\diff h'_{\mu\nu}}
\right)_{h' = 0}  \nonumber \\ = -
Z(g,F,T_0,U) \, \sum^{\infty}_{m=1} {\rm Tr} \left[\frac{h}{g +
2\pi\alpha' F + \frac{\alpha'}{2} \frac{U}{m}}\right],
\eeqn
where the trace is taken over the $\mu$ and $\nu$ indices.

\noindent
{\bf 3.}  We can compare this computation with a naive application of
the boundary state formalism.

The boundary state $|B \rangle$ is a quantum state of closed string theory
which obeys the boundary condition
\eq{boundary}:

\beq\label{boundary1}
\left(g_{\mu\nu} \, \diff_n \hat{X}^\nu(\theta) + 2\pi\alpha' F_{\mu\nu} \diff_\theta X^\nu (\theta) +
\frac{\alpha'}{2} U_{\mu\nu} \hat{X}^\nu(\theta)\right)|B\rangle = 0,
\eeq
where $\hat{X}(\theta)$ is the operator corresponding to the boundary
value of the map $X$ with the following closed string mode expansion:

\beq
\hat{X}^\mu(z,\bar{z}) = x^\mu_0 + p^\mu \log{z} +
\sum_{n\neq 0} \left[\frac{\alpha^\mu_n}{n} z^n +
\frac{\tilde{\alpha}^\mu_n}{n} \bar{z}^n\right].
\eeq
In this formula $z = e^{-\sigma_1 - {\rm i} \, \sigma_2} = \rho \,
e^{-{\rm i} \, \theta}$, $0 \le \rho \le 1$ is the complex coordinate
on the disc, $x_0$ and $p$ are
coordinate and momentum of the string center of mass, the sum runs
over $n$ from minus infinity to plus infinity except zero and the
generators $\alpha$ and $\tilde{\alpha}$ obey certain conditions to
make $\hat{X}$ hermitian, as well they obey the standard commutation
relations (see e.g., \cite{DiVecchia:1999fx,DiVecchia:1999rh}).

The solution to \eq{boundary1} is

\beq\label{state11}
|B\rangle = \cN \prod_{n \ge 1} \, \exp\left\{- \left[\frac{g - 2\pi \alpha' F - \frac{\alpha'}{2} \, \frac{U}{n}}{g +
2\pi \alpha' F + \frac{\alpha'}{2} \,
\frac{U}{n}}\right]_{\mu\nu} \, \frac{\alpha^\mu_{-n} \, \tilde{\alpha}^\nu_{-n}}{n}\right\}|0\rangle,
\eeq
where $|0\rangle$ is the vacuum state, which is annihilated by all
creation operators $\alpha_n$ and $\tilde{\alpha}_n$ with $n > 0$,
$\cN$ is a normalization constant.

The normalization is fixed by considering the coupling of the
off-shell (momentum zero) closed string tachyon whose coupling to the
boundary state should be equal to a trivial perturbation of the sigma
model partition function,

\beq\label{leg}
\cN = \langle 0 |\, B\rangle = Z(g,F,T_0,U)
\eeq

Now we would like to find (along the lines of
\cite{DiVecchia:1999fx,DiVecchia:1999rh}) the reaction of the
background closed string fields on the state $|B\rangle$. For this we
consider the correlator:

\beq\label{corr1}
\langle h |\, B\rangle = \langle 0| \alpha_1^\mu \tilde{\alpha}_1^\nu h_{\mu\nu} |B\rangle,
\eeq
which should be compared with the correlator \eq{corr}. However,
we obtain

\beqn\label{answer1}
\langle h |\, B\rangle = - Z(g,F,T_0,U) \, \left[\frac{g - 2\pi\alpha' F -
\frac{\alpha'}{2} \, U}{g + 2\pi\alpha' F + \frac{\alpha'}{2} \, U}\right]^{\mu\nu}
\, h_{\mu\nu}  \nonumber \\ = - 2 \, Z(g,F,T_0,U) \, {\rm Tr}
\left[\frac{h}{g + 2\pi\alpha' F + \frac{\alpha'}{2} \, U} \right],
\eeqn
where at the last step we used the fact that $h$ is
traceless.

The formula \eq{answer} clearly does not agree with~\eq{answer1}.
Note that they would agree if $U=0,\infty$~\cite{DiVecchia:1999fx,DiVecchia:1999rh}.
In fact, these
two expressions agree up to a (infinite) normalization factor at the
fixed points of the renormalization group flow, $U\to 0$ and
$U\to\infty$.  As we will see below the infinite factor will turn out
to be the volume of the non-compact group PSL(2,R)~~\cite{Liu:1988nz}.

\noindent
{\bf 4.}  The apparent paradox that we have arrived at should not be
surprising.  The application of the boundary state formalism to the
computation of the expectation value of a closed string vertex
operator in open string theory requires a conformal mapping of the
punctured disc, which is the world-sheet of open strings, to the
semi-infinite cylinder, which is the world-sheet of closed strings.
In the conformally non-invariant theory that we are considering here,
it is natural to expect that this mapping is blocked by the conformal
anomaly.

The global conformal group of the disc is PSL(2,R)~\footnote{The
relevance of PSL(2,R) in a similar context was previously noticed
in~\cite{Shatashvili:1993kk} and in~\cite{Craps:2001jp}.}.  If the
PSL(2,R) symmetry were {\it not} broken it would be possible to use it
to fix the position one point on the disc and one point on its
boundary (or three points on the boundary).  This could be used to get
rid of integration over $\sigma$ in~\eq{corr}. This means that:

\beq\label{correq}
\left\langle \int_D d^2\sigma h_{\mu\nu} \diff^a X^\mu(\sigma)
\diff^a X^\nu (\sigma) \right\rangle_{F,T_0,U = 0} =  \pi
\left\langle h_{\mu\nu} \diff^a X^\mu(\sigma') \diff^a X^\nu
(\sigma') \right\rangle_{F,T_0,U = 0}
\eeq
Here $\sigma'$ is some particular point on the disc, say $0$.
We expect that this will occur when $U=0$ or $U=\infty$.
However, since the conformal symmetry is broken when $U$ does not
have these values, the matrix element depends on the position
and the integration is important.

The conformal mapping of a point $z$ on the disc to a point $\eta$ in
the cylinder is $z=e^{-\eta}$.  In this mapping, the center of the disc,
at point $z=0$ is mapped to the cap of the cylinder at infinity, ${\rm Re}
\eta=\infty$.  In the boundary state computation which leads to \eq{corr1}
it is assumed that the boundary state is at one cap of the cylinder, where
${\rm Re}\eta=0$ and the graviton $\vert h\rangle$ is at the other cap which
is located at ${\rm Re}\eta=\infty$, which is the image of the center
of the disc.  For this reason, we expect the boundary state computation
to produce the expectation value of the graviton vertex operator inserted
at the center of the disc.

In the sigma model, it is straightforward to compute the correlator:

\beq
\langle h_{\mu\nu} \diff^a X^\mu \,
\diff^a X^\nu(\rho,\theta)\rangle_{F,T_0,U}.
\eeq
by summing the perturbation expansion for $U$, similar to computations
in refs.~\cite{Fradkin:1985qd,Laidlaw:2001kb}. In the course of the
calculation we use the boundary-to-disc propagator with Neuman
boundary conditions:

\beq
G(\rho \, e^{{\rm i} \, \theta}, \, e^{{\rm i} \, \theta'}) =
\frac{1}{\pi} \, \sum^{\infty}_{m=1} \frac{\rho^m}{m} \, \cos{[m \,
(\theta - \theta')]}
\eeq
which also gives the boundary-to-boundary propagator in the limit $\rho
\rightarrow 1$.  Explicitly the contribution to the correlator from $n$
interactions with the background $U$ is:

\beqn
Z(g,F,T_0,U) {\rm Tr} \left( \frac{- \alpha' U}{2} \right)^n
\times \nonumber \\ \times \frac{1}{\pi^n}
\int d\phi_1 \ldots d\phi_n \left(\partial_\rho \hat \rho +\frac{1}{\rho}
\partial_\phi \hat \phi \right) \sum_{m=1}^\infty
\frac{\rho^m \cos[ m (\phi-\phi_1) ]}
{m}  \nonumber \\ \times
\sum_{m_1=1}^\infty \frac{ \cos[ m_1 (
\phi_1 - \phi_2 )]}{m_1} \ldots \left(\partial_\rho \hat \rho
+ \frac{1}{\rho} \partial_\phi \hat \phi \right)
\sum_{m_n=1}^\infty \frac{\rho^{m_n}\cos[ m_n (\phi_n - \phi)]}{m_n}
\eeqn
The above integral is trivial, and upon summation over all values
of $n$ and inclusion of the antisymmetric field $F$ the result is:

\beqn\label{express}
\langle h_{\mu\nu} \diff^a X^\mu \, \diff^a X^\nu(\rho,\theta)\rangle_{F,T_0,U} =  \nonumber \\ 
- 2 \, Z(g,F,T_0,U)\, \sum^{\infty}_{m=1} m \, {\rm Tr}\left(\frac{h}{g + 2\pi \alpha' F + \frac{\alpha'}{2}\, \frac{U}{m}}\right) \rho^{2(m-1)}.
\eeqn
Now it is easy to see that to obtain \eq{answer} one has to integrate
this expression with the measure $\int \rho \, d\rho \, d\theta$,
while to obtain \eq{answer1} it is necessary to put $\rho = 0$: only
the $m=1$ term survives in this case.  This is in agreement with our
expectation that the boundary state describes the matrix element only
when the operator is inserted at the center of the disc.

\noindent
{\bf 5.}  With the above choice of coordinates on the disc, there is a
subset of the full conformal group of the plane which preserves the
position and shape of the boundary of the disc.  This subset is a
PSL(2,R) subgroup of the full conformal group.  It acts on the complex
coordinates of the disc as

\begin{equation}\label{transformation}
z\to w(z) = \frac{ az+b}{ b^* z +a^*},
\end{equation}
where
\begin{equation}
\vert a\vert^2-\vert b\vert^2=1.
\end{equation}
It is easy to verify that the unit circle is mapped onto itself.
Thus, this mapping preserves the boundary of the disc.  The origin is
mapped to the point $b/a^* = \rho \, e^{-{\rm i} \, \theta}$ in the
interior of the disc.

We will examine how the boundary state $|B\rangle$ behaves under this
transformation.

The boundary state is created by the exponential of the operator
\begin{equation}
{\cal B}=\sum_{n=1}^\infty S_{\mu\nu}(n)
\alpha^\mu_{-n}\tilde \alpha^\nu_{-n}, \quad {\rm and} \quad
S_{\mu\nu}(n) =
\frac{1}{n}\left[\frac{g - 2\pi \alpha' F - \frac{U}{n}}{g + 2\pi\alpha' F + \frac{U}{n}}\right]_{\mu\nu}
\label{bee}
\end{equation}
where $\alpha_n^\mu$ and $\tilde\alpha_n^\mu$ are closed string oscillators.
It is useful to write this operator in terms of position variables.
For this, we introduce the two fields,

\begin{eqnarray}
A^\mu(z)=\sum_{n\neq0}\frac{\alpha^\mu_{-n}}{n}z^n
\\
\tilde A^\mu(\bar z)=
\sum_{n\neq 0}\frac{\tilde \alpha^\mu_{-n}}{n}{\bar z}^n.
\end{eqnarray}
Here $z$ are complex coordinates on the disc and $\eta=-\ln z$ are
coordinates on a cylinder.

Then (\ref{bee}) can be written in the form

\begin{equation}
{\cal B}=\oint \frac{dz}{2\pi i} \oint\frac{d\bar z}{-2\pi i}
S_{\mu\nu}(z,\bar z)
A^\mu(z)\tilde A^\nu(\bar z),
\end{equation}
where the integrations are on the unit circle and the kernel is
defined by the power series

\begin{equation}
S_{\mu\nu}(z,\bar z)= \sum_{p=1}^\infty S_{\mu\nu}(p)~p^2~(z\bar z)^{p-1}.
\end{equation}
Now, we take into account that, under a general coordinate
transformation, the coordinate functions $A^\mu(z)$ and $\tilde A(z)$
transform like,

\begin{equation}
A'^\mu(z')=A^\mu(z), \quad
{\tilde A}'^\mu (\bar z')= {\tilde A}^\mu(\bar z).
\label{fieldtran}
\end{equation}
If we apply this equation to the conformal transformation and change
variables in the integral, we obtain the transformed boundary operator:

\begin{equation}
{\cal B}(a,b) =\oint \frac{dz}{2\pi i} \oint\frac{d\bar z}{-2\pi i}
S_{\mu\nu}(z,\bar z|a,b)
A^\mu(z)\tilde A^\nu(\bar z),
\label{boptran}
\end{equation}
where

\begin{equation}
S_{\mu\nu}(z,\bar z|a,b)= \left| \frac{dw}{dz}\right|^2
S_{\mu\nu}(w(z),\bar w(\bar z)).
\label{kertran}
\end{equation}

As an exercise, we can verify that the usual, conformally invariant
boundary state would be independent of the PSL(2,R) coordinates
$(a,b)$.  In that case (for simplicity we put here $F=0$),

\begin{equation}
S^0_{\mu\nu}(p)=\frac{\delta_{\mu\nu} }{p}
\end{equation}
and

\begin{equation}
S^0_{\mu\nu}(z,\bar z)=\frac{\delta_{\mu\nu}}{\left( 1- z\bar z\right)^2 }
\end{equation}
and

$$
S^0_{\mu\nu}(z,\bar z|a,b)= \delta_{\mu\nu} \left| \frac{dw(z)}{dz}\right|^2
\frac{ \delta_{\mu\nu} }{ \left( 1 - w(z)\bar w(\bar z)\right)^2 }
=\frac{\delta_{\mu\nu} }{ \left( 1-z\bar z\right)^2 }. $$ This is
independent of $a$ and $b$, which is the desired result.

Then, the PSL(2,R) transformed boundary state is created by the
exponential of the operator ${\cal B}(a,b)$. In terms of oscillators,
this operator has the form

\begin{equation}
{\cal B}(a,b)=\sum_{m,n>0} S_{\mu\nu}(m,n|a,b) \alpha^\mu_{-m}
\tilde \alpha^\nu_{-n}.
\label{oscker}
\end{equation}
(We will verify that it still contains only negative index
oscillators.)  The moments are defined by

\begin{equation}
S_{\mu\nu}(m,n|a,b)= \sum_{p=1}^\infty \frac{p}{m}\frac{p}{n}S_{\mu\nu}(p)
\oint \frac{dz}{2\pi i}z^{p-1} \left( \frac{ b^*z+a^*}{az+b}\right)^m
 \oint \frac{ d\bar z}{-2\pi i}{\bar z}^{p-1}
\left( \frac{ b\bar z+a}{a^*\bar z+b^*}\right)^n.
\label{ker}
\end{equation}
Since $|a|/|b|>1$, the contour integrals on the right-hand-side of
(\ref{ker}) have poles inside the unit circle only when $m,n>0$.
Therefore they are non-zero only when $m>0$ and $n>0$, as anticipated
in (\ref{oscker}).  It is straightforward to evaluate the integrals in
(\ref{ker}).  For example, in the case which we will see shortly is
relevant to massless closed string states,
\begin{equation}
S_{\mu\nu}(1,1|a,b)=\frac{1}{|a|^2} \sum_{p=1}^\infty
\left| \frac{b}{a} \right|^{2p-2} ~p^2 ~ S_{\mu\nu}(p)~.
\label{11}
\end{equation}
We can see from the form of the transformation in (\ref{ker}) that the
boundary states generally depend on all three parameters of PSL(2,R).
In some special cases, (\ref{11}) for example, it depends on fewer
parameters, such as $|b/a^*| = \rho$. The matrix element of any
massless closed string state with the boundary state will depend on
the PSL(2,R) parameters only through this dependence on the
coordinate.

Note that the matrix element $\langle 0|\,B\rangle$  does
not change under the transformation \eq{transformation}, i.e. the
eq. \eq{leg} is legitimate.  However, the correlator $\langle h|\,
B\rangle$ transforms according to \eq{11} as:

\beqn\label{answer2}
\langle h |\, B_\rho\rangle = \langle 0|
h_{\mu\nu} \alpha_1^\mu \tilde{\alpha}_1^\nu
\, Z(g,F,T_0,U) \times \nonumber \\ \times
\exp\left\{- \sum_{m>0} \frac{1}{m} \,
\left[\frac{g - 2\pi \alpha' F  - \frac{\alpha' U}{2 m}}{g +
2\pi\alpha' F + \frac{\alpha' U}{2 m}}\right]_{\mu\nu} m^2
(1-\rho^2)^2 (-\rho)^{2(m-1)} \alpha^\mu_{-1} \,
\tilde{\alpha}^\nu_{-1} + ... \right\}|0\rangle =
\nonumber \\ = - 2 Z(g,F,T_0,U) \, \sum_m m
{\rm Tr}\left[\frac{h}{g + 2\pi\alpha' F + \frac{\alpha' U}{2 m}}
\right] \, \rho^{2(m-1)}\, (1 - \rho^2)^2.
\eeqn
It is worth mentioning here that if $U\to 0$ then $\sum_{m>0} m
\rho^{2(m-1)} = \frac{1}{(1-\rho^2)^2}$ exactly cancels $(1-\rho^2)^2$
in the numerator and, hence, \eq{answer} agrees with \eq{answer2},
\eq{answer1}.

At the same time the Haar measure on the PSL(2,R) group is given by:

\beqn\label{mes}
\int d^2 a d^2 b \, \delta(|a|^2 - |b|^2 - 1) f = \nonumber \\ =
\int d^2 a d^2 b \, \delta(|a|^2 - |b|^2 - 1) \int d\rho \,
\delta\left(\left|\frac{b}{a}\right| - \rho\right) f =
2 \, \pi^2 \, \int \frac{\rho \, d\rho}{(1 - \rho^2)^2} f,
\eeqn
which is valid if the function $f$ within the integral depends only on
$\rho$.  Combining the formulas \eq{answer2} and \eq{mes} we find
exact agreement.

\noindent
{\bf 6.}  In conclusion, we conjecture that, the average over PSL(2,R)
of the transformed boundary state,

\beq
|\hat B\rangle=
\int  d^2 a d^2 b \, \delta(|a|^2 - |b|^2 - 1) \, |B_{a,b}\rangle,
\eeq
will have the correct overlap with any on-shell closed string state.
Here we have checked this for the closed string tachyon, the graviton
and it is straightforward to check it for the anti-symmetric tensor
which has a non-zero expectation value when a background gauge field
is turned on.  It would be interesting to check this hypothesis for
higher order correlation functions. The generalization of our results to the superstring boundary
states with linear tachyon profile is straightforward.

\subsection*{Acknowledgments}

We are grateful to Taejin Lee for many discussions in the early part of
this work. This work was supported by NSERC of Canada and
NATO Collaborative Linkage Grant SA(PST.CLG.977361)5941.
The work of E.T.A.  was also supported in part by
NSERC NATO Science Fellowship and RFFI 01-02-17488.

\end{document}